\documentclass[11pt,a4paper]{article} 
\usepackage{jheppub} 

\usepackage{graphicx}
\usepackage{color}
\usepackage{dcolumn}

\newcommand {\beq}{\begin{eqnarray}}
\newcommand {\eeq}{\end{eqnarray}}

\title{Quark-anti-quark potentials from Nambu-Bethe-Salpeter amplitudes on lattice}

\author[a,b,1]{Yoichi Ikeda
\note{Email : yikeda@riken.jp}}
\author[b,2]{and Hideaki Iida
\note{Email : hiida@riken.jp}}
\affiliation[a]{Department of Physics, 
Tokyo Institute of Technology, Meguro, Tokyo 152-8551, Japan}
\affiliation[b]{RIKEN Nishina Center, 2-1, Hirosawa, Wako, Saitama 351-0198, Japan}

\abstract{

Quark--anti-quark ($\bar q$-$q$) potentials with finite quark masses are studied from 
the $\bar q$-$q$ Nambu-Bethe-Salpeter (NBS) 
wave functions in quenched lattice QCD. 
With the use of a method which has been recently developed 
in the derivation of nuclear forces from lattice QCD,
we derive the $\bar q$-$q$ potentials from the NBS wave functions.
We calculate the $\bar q$-$q$ NBS wave functions 
in pseudo-scalar and vector channels for several quark masses.
The derived potentials at each quark mass in both channels  
show linear plus Coulomb form. 
We also discuss the quark-mass and channel dependence of 
the $\bar q$-$q$ potentials. 
}

\begin{document}

\maketitle


\section{Introduction}

An inter-quark potential is one of the most important ingredients 
of quantum chromodynamics (QCD).
Experimentally, Regge slope \cite{Chiu1972a} suggests that 
the inter-quark potentials show linear behavior at long distance.
The string tension of 
the potentials between a quark and an anti-quark,
$\sigma\simeq 1.3{\rm GeV}$, can be roughly estimated by 
hadron spectra using the relation $J=M^2/(4\sigma)$ with the spin $J$ and 
the mass $M$ of hadrons.
At short distance, the inter-quark potential shows 
like the Coulomb interaction, which is, for example,
suggested by the analogy between 
quarkonium and positronium.
In fact, the linear plus Coulomb behaviors of the inter-quark potentials 
reproduce the low-lying hadron spectra well in quark models. 

Theoretically, 
the study of the inter-quark potentials is 
challenging issue due to 
the non-perturbative nature of low energy phenomena in QCD.
Lattice QCD simulation is the powerful tool for a numerical
investigation in such a strong-coupling region of QCD. 
From the expectation value of Wilson loops,
the potential for an infinitely heavy quark 
and  anti-quark ($\bar Q$-$Q$ potential) and also 
the three-quark potential ($3Q$ potential)
can be obtained on lattices~\cite{Bali2001,TTT2000}.
The $\bar Q$-$Q$ potential from quenched lattice QCD simulations reveals
the form of $V(r)=\sigma r - A/r$
with $\sigma=0.89{\rm GeV/fm}$ and $A=0.26$. 

The actual inter-quark potentials suffer from the effect of quark motions,
which is not included in the $\bar Q$-$Q$ potential. 
One can take into account the corrections coming from finite quark masses $m_q$ 
order by order with the use of the heavy quark effective field theory. 
The effective field theory utilizes the hierarchy of scales coming 
from the heavy quark mass $m_q$ and the relative velocity of heavy quarks, $v$. 
The potential nonrelativistic QCD (pNRQCD) is such an effective field 
theory at the ultrasoft scale $m_q v^2$ obtained by integrating out 
the hard scale $m_q$ and the soft scale $m_q v$ 
~\cite{Bali2001,Brambilla,Brown1979a,Eichten1979,Koma2007a}. 
It is convenient to employ pNRQCD to
obtain the corrections of heavy quarkonium spectra 
to the heavy quark motion.

In this work, we study potentials between 
a quark and an anti-quark with a finite mass ($\bar q$-$q$ potentials) 
from quenched lattice QCD simulations. 
In order to explore the $\bar q$-$q$ potentials,
we apply the systematic method 
which utilize the equal-time Nambu-Bethe-Salpeter (NBS) amplitudes
to extract hadronic potentials
~\cite{Ishii2007a,Aoki2010,Nemura2008,Nemura2009,Murano2011,Inoue2010a,Inoue2010b,
Doi2010,Sasaki2010,Ikeda2010,Kawanai2010,Hatsuda2011, TTT2009}  
to the systems with a relatively light quark and an anti-quark.
Due to the absence of the asymptotic fields of quarks, 
the reduction formula cannot be applied directly. 
Therefore, we assume that the equal-time NBS amplitudes for the $\bar q$-$q$ systems
satisfy the Nambu-Bethe-Salpeter (NBS) equation with constant quark masses
which could be considered as the constituent quark masses.
By using the derivation of the relativistic three-dimensional formalism from
the NBS equation developed by L\'{e}vy, Klein and Macke (LKM formalism)
~\cite{Levy1952,Klein1953,Klein1974,Macke1953},
we shall obtain the $\bar q$-$q$ potentials 
{\it without expansion in terms of $m_q$}.
The preliminary results of the NBS wave functions and potentials of the $\bar q$-$q$
systems have been reported in Ref.~\cite{IkeIida2010}.

The paper is organized as follows.
In Sec.~2, we present our method to extract the $\bar q$-$q$ potentials.  
In Sec.~3, we show the lattice QCD setup. 
We then show our numerical results of the $\bar q$-$q$ wave functions and potentials in 
pseudo-scalar and vector channels for four different quark masses in Sec.~4. 
The obtained potentials reveal the linear plus Coulomb forms 
which are similar to the $\bar Q$-$Q$ potential from the Wilson loop. 
We perform fitting analyses of the $\bar q$-$q$ potential data.  
Sec.~5 is devoted to discussions and summary. 

\section{Method of the extraction of inter-quark potentials}

Following the formulation to 
define the potentials on lattices~\cite{Ishii2007a,Aoki2010,Klein1974},
we show the basic equations to extract the $\bar q$-$q$ potentials 
on the lattice below.
As shown in Ref. \cite{Klein1974}, 
the equal-time choice of the Nambu-Bethe-Salpeter (NBS) amplitudes satisfy 
the relativistic Schr\"{o}dinger-type equation 
{\it without an instantaneous approximation} for original interaction kernels of the NBS equation.
Therefore, we can start with the Schr\"{o}dinger-type equation 
(which is reffered as LKM equation in Ref \cite{Klein1974})
for the NBS wave function $\phi(\vec{r})$ 
to define potentials:
\begin{equation}
-\frac{\nabla^2}{2\mu} \phi(\vec{r}) + \int d\vec{r}' 
U(\vec{r},\vec{r'}) \phi(\vec{r'})
= E \phi(\vec{r}),
\label{eff-Sch-eq}
\end{equation}
where $\mu(=m_q /2)$ and $E$ denote the reduced mass of the $\bar q$-$q$ system
and the non-relativistic energy, respectively,
and we simply assume nonrelativistic kinematics.
Note that the potential $U(\vec r, \vec r')$ is generally energy-independent and non-local~\cite{Aoki2010,Kroli1956}.
In appendix A, we discuss  potentials derived from 
the relativistic Schr\"{o}dinger-type equation.
The relativistic effects may be necessary to reproduce the meson mass spectra
in the wide energy region.

For the two-nucleon case, it is proved that the Schr\"{o}dinger-type
equation is derived by using the reduction formula~\cite{Aoki2010}. 
Due to the absence of asymptotic fields for confined quarks, 
we suppose that
the $\bar q$-$q$ systems satisfy the NBS equation with their 
constant quark masses. 
In this study, constant quark masses $m_q$ are determined by 
half of vector meson masses $M_V$, i.e., $m_q=M_V/2$, 
as usually taken in constituent quark models.
Then, one finds Schr\"{o}dinger-type equation of Eq. (\ref{eff-Sch-eq}) 
as a three-dimensional reduction of NBS equation
by applying 
LKM~\cite{Levy1952,Klein1953,Klein1974,Macke1953} method. 

The energy-independent and non-local potential $U(\vec{r},\vec{r'})$ 
can be expanded in powers of the relative velocity 
$\vec{v}=-i\nabla/\mu$ of $\bar q$-$q$ systems at low energies,
\begin{eqnarray}
U(\vec{r},\vec{r'}) & = &
V(\vec{r}, \vec{v}) \delta(\vec{r} - \vec{r'})  \nonumber \\
& = & (V_{LO}(\vec{r})+V_{NLO}(\vec{r})+ \cdots) \delta(\vec{r}-\vec{r'}),
\label{exp}
\end{eqnarray}
with
\begin{eqnarray}
V_{LO}(\vec{r}) &=&
V_C(r)+V_T(r)S_{12},  \\
V_{NLO}(\vec{r}) &=&
V_{LS}(r) \vec L \cdot \vec S ,
\end{eqnarray}
where the $N^n LO$ term is of order $O(\vec{v}^n)$, and
$S_{12}$, $\vec L$ and $\vec S$ being the tensor operator,
orbital angular momentum and spin of the $\bar q$-$q$ systems, respectively.
Note that 
the velocity expantion is different from the usual $1/m_q$ expansion,
and
the central force  $V_C(r) $ in the leading order
includes not only linear and Coulomb terms 
but also higher order terms in the $1/m_q$ expansion 
such as the spin-spin interaction,
$V_{\rm spin}(r) \vec{\sigma}_{\bar q} \cdot \vec{\sigma}_{q}$,
which is regarded as an order $O(1/m_q^2)$.
This spin-spin interaction is an important ingredient of the mass formula
in the constituent quark model.
At the leading order, one finds
\begin{equation}
V(\vec{r}) \simeq V_{LO}(\vec{r}) = 
\frac{1}{2\mu}\frac{\nabla^2 \phi(\vec r)}{\phi(\vec{r})}+E.
\label{pot}
\end{equation}
The effective leading order $\bar q$-$q$ potentials $V_{LO}(\vec r)$ 
are studied in this work.
The convergence of the expansion of $\vec v$ can be checked by studying the 
energy dependence of the local potential $V_{LO}(\vec r)$ 
as in Ref.~\cite{Murano2011}.  
When the local potential $V_{LO}(\vec r)$ has little energy dependence, 
the potential between $\bar q$-$q$ is well described only by $V_{LO}(\vec r)$. 
In contrast, if the energy dependence is large, 
higher order terms are necessary. 
The study of the energy dependence of the local potential 
in the $\bar q$-$q$ systems is 
an important future work. 

In order to obtain the NBS wave functions of the $\bar q$-$q$ systems 
on the lattice,
let us consider the following equal-time NBS amplitudes 
\begin{align}
&\chi (\vec x + \vec r, \vec x, t-t_0; J^{\pi}) \nonumber \\
&=\left\langle 0 \right|
\bar{q}(\vec x + \vec r, t) \Gamma q(\vec x, t)
\overline{{\cal J}}_{\bar qq}(t_0; J^{\pi})
\left| 0 \right\rangle \nonumber \\
 &= 
\sum_{n}A_n \left\langle 0 \right|
\bar{q}(\vec x+\vec r, t) \Gamma q(\vec x, t)
\left| n \right\rangle 
\ e^{-M_n(t-t_0)},
\label{4-point}
\end{align}
with the matrix elements
\begin{equation}
A_n = \left\langle n \right| 
\overline{{\cal J}}_{\bar qq}(t_0; J^{\pi})
\left| 0 \right\rangle.
\end{equation}
Here $\Gamma$ represents the Dirac $\gamma$-matrices,
and $\overline{{\cal J}}_{\bar qq}(t_0; J^{\pi})$ stands for the source term
which creates the $\bar q$-$q$ systems 
with spin-parity $J^{\pi}$ on the lattice.
The NBS amplitudes in Eq. (\ref{4-point})
are dominated by the lowest mass state of mesons with the mass $M_0$
at large time separation ($t \gg t_0 $):
\begin{eqnarray}
\chi(\vec r, t-t_0; J^{\pi}) 
&=&
\frac{1}{V}
\sum_{\vec x}
\chi (\vec x+\vec r, \vec x, t-t_0; J^{\pi}) \nonumber \\
&\rightarrow&
A_0 \phi (\vec r; J^{\pi}) e^{-M_0(t-t_0)},
\label{4-point2}
\end{eqnarray}
with $V$ being the volume of the box.
Thus, the $\bar q$-$q$ NBS wave function is defined by the spatial correlation of
the NBS amplitudes.

The NBS wave functions in S-wave states are obtained under 
the projection onto zero angular momentum ($P^{(l=0)}$),
\begin{equation}
\phi(\vec r; J^\pi) =
\frac{1}{24}\sum_{g \in O} P^{(l=0)} 
\phi (g^{-1} \vec r; J^\pi),
\label{BS-wave}
\end{equation}
where $g \in O$ represents 24 elements of the cubic rotational group,
and the summation is taken for all these elements.
Using Eq. (\ref{pot}) and Eq. ({\ref{BS-wave}}),
we will find the $\bar q$-$q$ potentials and 
NBS wave functions from lattice QCD.

\begin{table}[htbp]
   \centering
   \begin{tabular}{cccccc}
      \hline
      \hline
      $\beta$ & $a$ & lattice size & volume &  $N_{\rm conf}$\\
      \hline 
     6.0 & 0.104 {\rm fm}  & $32^3\times 48$& (3.3fm)$^3$ & 100 \\
      \hline
      \hline
   \end{tabular}
   \caption{Simulation parameters used in this work. Scale is set 
   by string tension \cite{TTT2000,Iritani2009}. }
   \label{tab1}
\end{table}

\section{Numerical setup of the lattice simulations}
\label{sec:3}

In this section,
we show the actions and simulation parameters in this work.
We employ the standard plaquette gauge action,
\begin{align}
S_G[U] \equiv \frac{\beta}{N_c}
\sum_{x,\mu,\nu} {\rm Re}{\rm Tr}\{1-P_{\mu\nu}(x)\},
\end{align}
with $\beta\equiv 2N_c/g^2$. The plaquette $P_{\mu\nu}$ is defined as
\begin{align}
P_{\mu\nu}=U_\mu(x)U_\nu(x+\hat\mu)U^\dagger_\mu(x+\hat \nu)U_\nu^\dagger(x),
\end{align}
where $U_{\mu}(x)$ is a link variable. 
As for quark fields $\psi(x)$,
we adopt the standard Wilson fermion action,
\begin{align}
&S_F[\bar{\psi},\psi,U]\equiv  \sum_{x,y}\bar \psi(x)K(x,y)\psi(y),\\
&K(x,y)\equiv   \nonumber \\
&\delta_{x,y} - \kappa\sum_\mu\{({\bf 1}-\gamma_\mu)U_\mu(x)\delta_{x+\hat\mu,y}
+({\bf 1}+\gamma_\mu)U_\mu^\dagger(y)\delta_{x,y+\hat\mu}\},
\end{align}
where $\kappa$ is the hopping parameter.

We generate the quenched gauge fields 
on a $32^3\times 48$ lattice 
with QCD coupling $\beta = 6.0$,
which corresponds to the physical volume $V=(3.3{\rm fm})^3$ and 
the lattice spacing $a=0.104$fm~\cite{TTT2000,Iritani2009}.
We measure the $\bar q$-$q$ NBS wave functions for 
four different hopping parameters
$\kappa=0.1520$, $0.1480$, $0.1420$, 
$0.1320$: 
the corresponding pseudo-scalar (PS) 
meson masses $M_{\rm PS}$
in the calculation are 0.94, 1.27, 1.77, 2.53GeV, 
and vector (V) meson masses $M_{\rm V}$=1.04, 1.35, 1.81, 2.55GeV,
respectively. 
The number of configurations 
used in this simulation is 100 for each quark mass.
The simulation parameters are summarized in Table 1. 
The calculation of the $\bar q$-$q$ NBS wave functions requires gauge fixing, 
because $q$ and $\bar q$ operators are spatially separated at the sink time slice.
Here we adopt Coulomb gauge, which is frequently used for 
studies of hadron spectroscopy in lattice QCD. 
As for the source operator of the $\bar q$-$q$ systems, 
we employ a static wall source in Eq. (\ref{4-point}),
\begin{equation}
\overline{{\cal J}}_{\bar qq}(t_0; J^{\pi})=\bar Q(t_0) \Gamma Q(t_0),
\end{equation}
with the static wall quark operator
\begin{equation}
Q(t_0) \equiv \sum_{\vec x} \psi(\vec x, t_0).
\end{equation}

We note that gauge fixings and sink operators can be arbitrary chosen 
in the formalism, 
and all potentials with different gauge fixings and sink operators give 
the same physical observables, i.e., mass spectra and scattering lengths 
for instance. 
In this work, we employ Coulomb gauge and 
a local operator for sink operators. 
We can take another gauge fixing and sink operator, 
and the potential obtained with these conditions is generally different from that 
obtained in this work. 
In appendix B, we discuss the sink-operator dependence of the potential 
by using a gauge-invariant smeared operator for the sink.



\section{Numerical results for the $\bar q$-$q$ potentials}
\label{sec:4}

\begin{figure}[t]
\includegraphics[width=0.45\textwidth,clip]{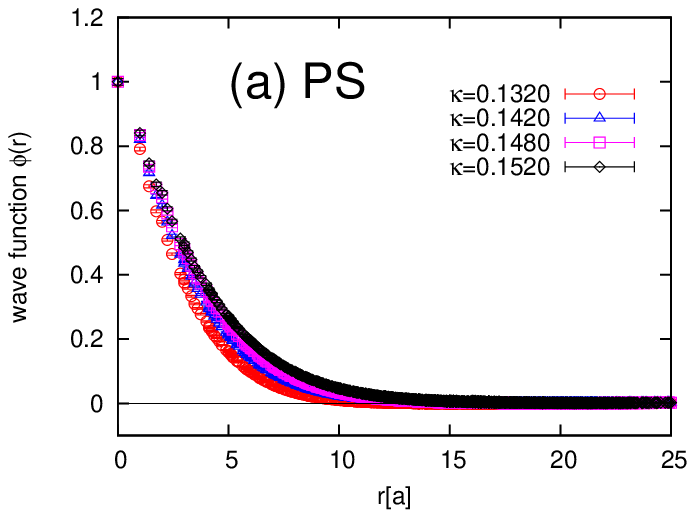}
\includegraphics[width=0.45\textwidth]{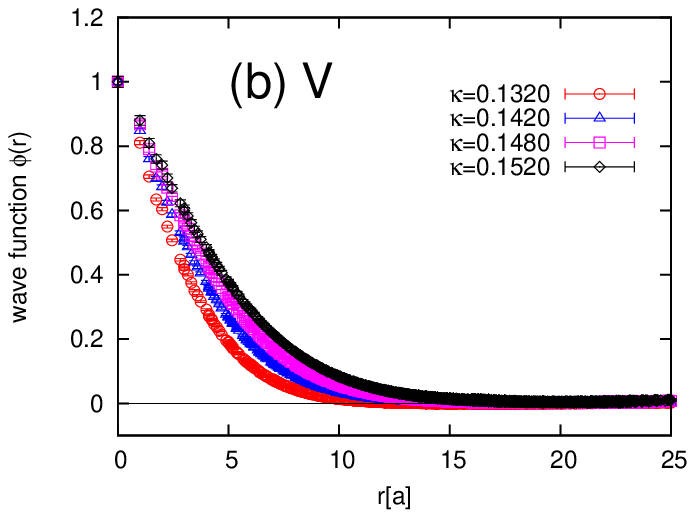}
\caption{
The $\bar q$-$q$ NBS wave functions in PS(a) and V(b) channels. 
The wave functions are normalized at origin. 
All the wave functions are localized in the box and indicate the bound states.
 }
\label{fig1}
\end{figure}

First, we show the numerical results of the NBS wave functions in Fig.~\ref{fig1}. 
Fig.~\ref{fig1}(a) and (b) are the NBS wave functions 
for each quark mass in PS and V channels, respectively, 
at the time slice $t=20$.
The NBS wave functions mostly vanish
at $r=1.5$fm for all quark masses in both channels. 
This indicates that
the spatial volume $V=(3.3{\rm fm})^3$ is large enough 
for the present calculations. 
The size of a wave function with a lighter quark mass becomes larger than 
that with a heavier one.
Comparing the results in PS and V channels, 
little channel dependence between PS and V channels is found, 
although the quark-mass dependence of the wave functions 
is a bit larger for V channel.  

\begin{figure}[t]
\includegraphics[width=0.45\textwidth]{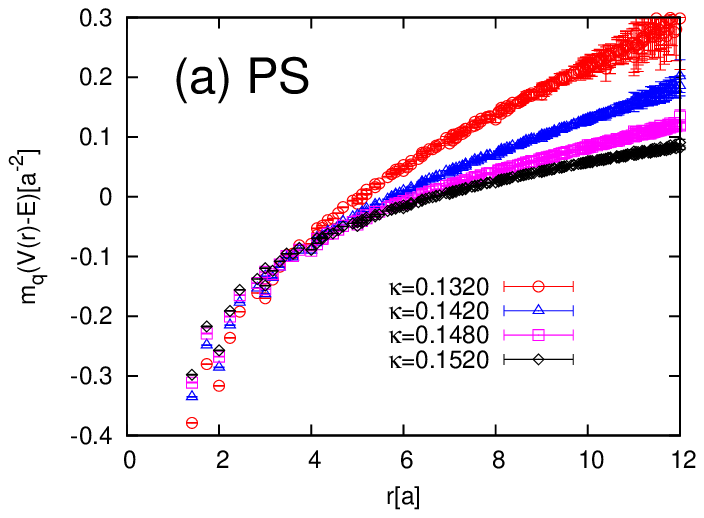}
\includegraphics[width=0.45\textwidth]{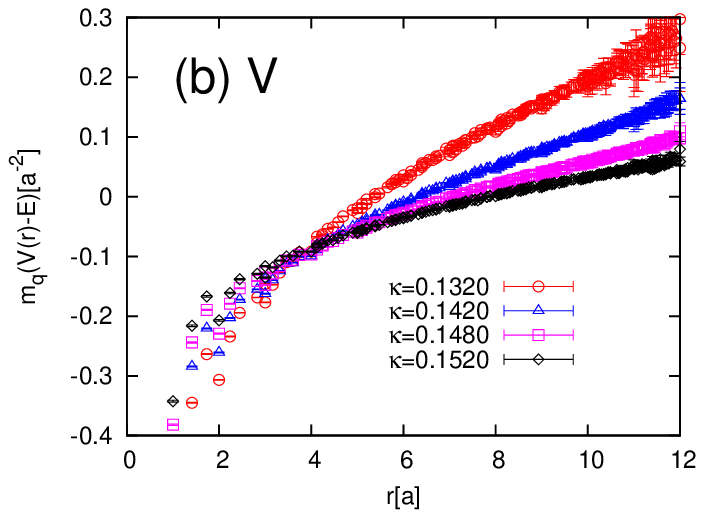}
\caption{Plots of $\nabla^2 \phi(r)/\phi(r) = 2\mu(V(r)-E)$  
in PS channel (a) and V channel (b) for 
each quark mass. 
The potentials show the linear plus Coulomb form. }
\label{fig2}
\end{figure}
\begin{figure}[t]
\includegraphics[width=0.45\textwidth]{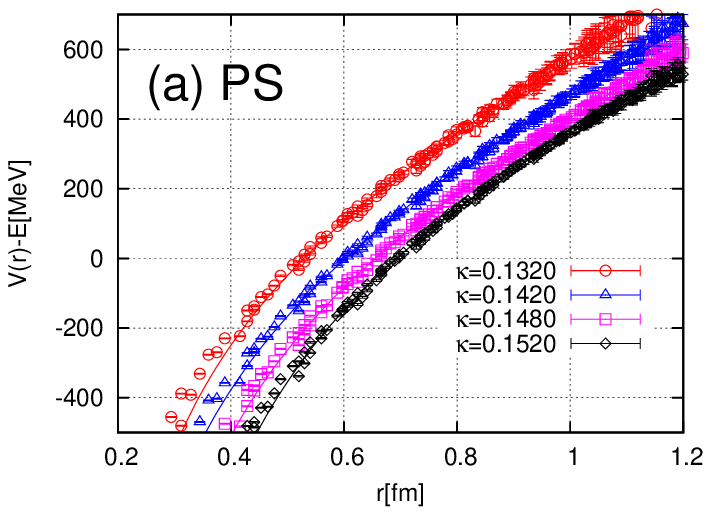}
\includegraphics[width=0.45\textwidth]{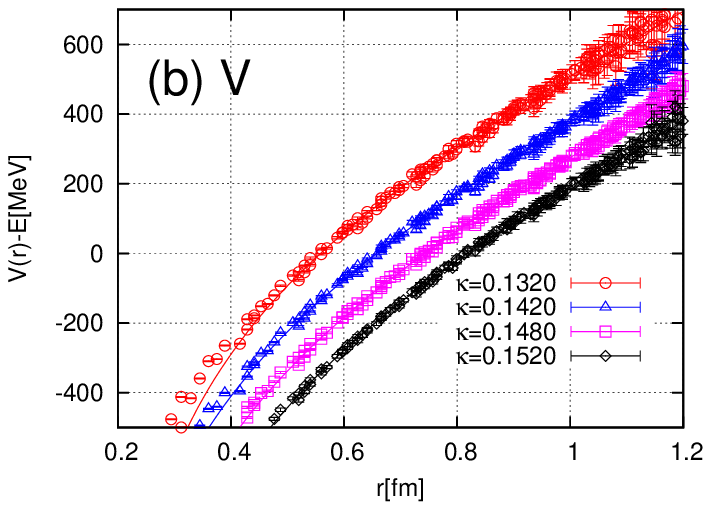}
\caption{Plots of the potential with
arbitrary constatnt energy shift $V(r)-E=\nabla^2 \phi(r)/(2\mu\phi(r))$   
in PS channel (a) and V channel (b) for 
each quark mass. 
The solid curves are the fit function with linear plus Coulomb form 
shown in Table \ref{tab2}. 
}
\label{fig3}
\end{figure}

In Fig.~\ref{fig2}, we show the Laplacian parts of $\bar q$-$q$ potentials in Eq.~(\ref{pot}),
$\nabla^2 \phi(r)/\phi(r)$, 
for each quark mass and channel.
Fig.~\ref{fig2}(a)  
shows $\nabla^2 \phi(r)/\phi(r)=2\mu(V(r)-E)$  
in PS channel for each quark mass at the time slice $t=20$. 
As shown in Fig.~\ref{fig2}, 
one can 
see that the potential form is similar to that obtained from Wilson loop, 
namely, the potential form looks like linear plus Coulomb form, 
although the derivation of the potentials 
is largely different between these two methods. 
Fig.~\ref{fig2}(b) represents  
$\nabla^2 \phi(r)/\phi(r)$ in V channel for each quark mass 
at the same time slice $t=20$. 
The basic properties are similar to that in PS channel, 
although quark mass dependence is a bit larger for V channel. 

Fig.~\ref{fig3}(a) and (b) 
show the potentials with arbitrary energy shifts $E$,
i.e, $V(r)-E$$=$$\nabla^2 \phi(r)/(2\mu \phi(r))$ 
in PS and V channels, respectively, 
for each quark mass at the time slice $t=20$. 
Note that the quark mass $m_q (= 2\mu)$ is determined 
by the half of vector meson mass, $m_q = M_V/2$,
as mentioned in the previous section.


We perform fit analyses of the potentials in Fig.~\ref{fig3}(a) and (b). 
For the fit function, we choose the linear plus Coulomb form, 
$f(r)=\sigma r - A/r + C$. 
We fit $f(r)$ to the potential data for each quark mass and channel.
We use the on-axis data with the range $3<r/a<10$ in the fit.
The fit results are summarized in Table \ref{tab2},
and denoted by solid curves in Fig.~\ref{fig3}(a) and (b).
$\chi^2/N_{\rm df}$ is around 0.5 for all the fit, 
which means the data are well described by the linear plus Coulomb form. 
We find moderate quark mass dependence of the string tension.
The string tension becomes larger as increasing quark masses in both channels, 
and that for heaviest quark mass in our simulation is about 820MeV/fm, which is comparable to  
that obtained from an expectation value of the Wilson loop. 
On the other hand, the Coulomb coefficient $A$ strongly depends on quark masses.
The Coulomb coefficient becomes small as increasing quark masses, 
and is roughly approaching to that obtained from an expectation value of the Wilson loop. 
\begin{table*}[hptb]
\begin{center}
\begin{tabular}{cccccccccc}
\hline \hline
\multicolumn{1}{c}{}& \multicolumn{3}{c}{Pseudo-scalar} & \multicolumn{3}{c}{Vector} \\
\hline
$\kappa$  & $\sigma(m_q,i)$ & $A(m_q,PS)$ &  $\chi^2/N_{\rm df}$ &  $\sigma(m_q,i)$ &
 $A(m_q,V)$ &  $\chi^2/N_{\rm df}$  \\
&MeV/fm & MeV$\cdot$fm &   & MeV/fm 
 & MeV$\cdot$fm & & \\
\hline
$0.1320$ & $ 819 (47)$ & $ 215 (7)$ &  0.32& $825 (48)$ & $ 195 (7)$ &  0.63\\
$0.1420$ & $ 753 (34)$ & $ 264 (5)$ &  0.35& $ 765(37)$ & $216 (6)$ &  0.61\\
$0.1480$ & $ 691 (30)$ & $ 338 (5)$ &  0.46& $ 723 (39)$ & $ 249(7)$ &  0.44 \\
$0.1520$ & $ 601 (29)$ & $ 443 (5)$ &  0.31& $ 697 (63)$ & $ 291 (13)$ &  0.23 \\
\hline \hline
\end{tabular}
\caption{
The fitting results of the potentials in Fig.~\ref{fig3}.
The function to be fitted is 
$f(r)=\sigma(m_q,i) r -A(m_q,i)/r +C(m_q,i)$. 
The fit range is $3< r/a <10$.
}
\label{tab2}
\end{center}
\end{table*}

Next, we perform another type of fit analyses. 
Assuming that the string tension $\sigma$ is independent of 
the quark masses due to the quenched QCD simulations, 
where the contributions from quark loops are eliminated, 
we perform the fit by minimizing the general $\chi^2/N_{\rm df}$ 
which is defined as $\chi^2/N_{\rm df}=\sum_{m_q , i}\chi^2(m_q, i)/N_{\rm df}$ 
with $i=PS,\ V$~\cite{Hohler1976}. 
We call the fit ``universal fit" here. 
The fit functions $f_1(r)$ can be explicitly written by 
$f_1(r)=\sigma r- A(m_q,i)/r + C(m_q,i)$. 
The free parameters of the fit are 
$\sigma, A(m_q,i),$ $C(m_q,i)$ for $f_1(r)$. 
In the fit, we choose the range of the potential data as 
$3 \le r/a \le 10$. 
The fitting results are shown in Table \ref{tab3}. 
The general $\chi^2/N_{\rm df}$ is achieved with $1.52$ for $f_1(r)$
with $\sigma = 723 \ (30)$ MeV/fm.  
Since our simulation includes all the quark mass effects, 
$f_1(r)$ is modified by the higher order effect of $1/m_q$ expansion. 
\begin{table*}[hptb]
\begin{center}
\begin{tabular}{ccccc}
\hline \hline
\multicolumn{1}{c}{}& \multicolumn{1}{c}{Pseudo-scalar} & \multicolumn{1}{c}{Vector} \\
\hline
$\kappa$  & $A(m_q,PS)$ &
 $A(m_q,V)$  \\
 & MeV$\cdot$fm 
 & MeV$\cdot$fm  \\
\hline
$0.1320$ & $     231.45   (4.50)  $ & $       212.42   (4.60)     $  \\
$0.1420$ & $     270.63    (4.53)   $ & $  224.91    (4.64)       $  \\
$0.1480$ & $      331.91    (4.95)      $ & $    248.96   (5.00)  $  \\
$0.1520$ & $     414.93   (5.89)        $ & $     284.76 (7.03)   $  \\
\hline \hline
\end{tabular}
\caption{
The fitting results of the potentials for $f_1(r)$ 
without the channel and the quark mass dependences of the string tension.
The obtained string tension is $\sigma = 723 (30)$ MeV$/$fm.
The fit range is $3< r/a < 10$. 
}
\label{tab3}
\end{center}
\end{table*}

We also perform another type of the universal fit with the fit function, 
$f_2(r)=\sigma r -A(m_q,i)/r +C(m_q,i)+B_1(m_q,i) \log(r)-B_2(m_q,i)/r^2$,  
which contains $1/m_q$ correction terms, i.e., $1/r^2$ and $\log (r)$. 
The $\log$ term is 
predicted by pNRQCD and the effective string theory, respectively~\cite{Koma2008,Soto2008}. 
In this fit, we use not only on-axis data but also off-axis data,
and results are
summarized in Table~\ref{tab4} ($\chi^2/N_{\rm df} = 2.14$).
As shown in Table~\ref{tab4}, 
if we employ the fit function $f_2(r)$ 
in which $O(1/m_q)$ terms are taken into account,
the Coulomb coefficients are smaller than those obtained $f_1(r)$ and are comparable to the values from Wilson loop. 

\begin{table*}[hptb]
\begin{center}
\begin{tabular}{ccccccccc}
\hline \hline
\multicolumn{1}{c}{}& \multicolumn{3}{c}{Pseudo-scalar} & \multicolumn{3}{c}{Vector} \\
\hline
$\kappa$  & $A(m_q,PS)$ & $B_1(m_q,PS)$ & $B_2(m_q,PS)$ &
 $A(m_q,V)$ & $B_1(m_q,V)$ & $B_2(m_q,V)$ \\
& MeV$\cdot$fm & MeV & MeV$\cdot$fm$^2$
& MeV$\cdot$fm & MeV & MeV$\cdot$fm$^2$ \\
\hline
    $0.1320$ & $93.6 (139)$ &  $161 (128)$ & $20.8 (43.4)$ &
 $103 (129)$ & $119 (128)$ & $21.2 (38.0)$ \\
    $0.1420$ & $90.5 (172)$ & $164 (91.9)$ & $26.0 (42.7)$ &
 $81.3 (123)$ & $173 (97.8)$ & $15.9 (34.8)$ \\
   $0.1480$ & $93.2 (142)$ & $123 (76.2)$ & $45.2 (35.8)$ &
 $101 (106)$ & $146 (97.2)$ & $19.2 (35.5)$ \\
   $0.1520$ & $91.0 (65.4)$ & $48.2 (112)$ & $78.8 (19.7)$ &
  $276 (267)$ & $1.66 (44.2)$ & $ 1.13 (78.2)$ \\
\hline \hline
\end{tabular}
\caption{
The fitting results of the potentials for the fitting function 
$f_2(r)=\sigma r -A(m_q,i)/r +C(m_q,i)+B_1(m_q,i) \log(r)-B_2(m_q,i)/r^2$. 
The obtained string tension is $\sigma = 705 (120)$ MeV$/$fm. 
The fit range are is $5 \le r/a \le 12$.
}
\label{tab4}
\end{center}
\end{table*}

\section{Discussion and summary}
\label{sec:5}

We have studied the inter-quark potentials between 
a quark and an anti-quark ($\bar q$-$q$ potentials)
from the $\bar q$-$q$ Nambu-Bethe-Salpeter (NBS) wave functions. 
For this purpose, we have utilized 
the method which has been recently developed in the 
calculation of nuclear force from QCD~\cite{Ishii2007a,Aoki2010}.
We have calculated the NBS wave functions 
for the $\bar q$-$q$ systems with four different quark masses in pseudo-scalar 
and vector channels
and obtained the $\bar q$-$q$ potentials
through the Sch\"{o}dinger-type equation.
In this framework, the $\bar q$-$q$ potentials basically contains
full quark motions with the finite masses.
As a result, we have found that the shapes of the $\bar q$-$q$ potentials are
the linear plus Coulomb form which is similar to
the static $\bar Q$-$Q$ potential obtained from Wilson loop.

For the fitting, 
we have employed two types of fitting functions.
One is the linear plus Coulomb form regarded as the leading order (LO) terms
in the $1/m_q$ expansion.
The other function includes the next leading order (NLO) terms in addition to LO terms.
The fitting results with LO terms reveal that the Coulomb coefficients 
depend on the quark masses and are larger than those predicted from Wilson loop. 
On the other hand, if we have employed the NLO terms
together with the LO terms, the Coulomb coefficients become smaller and are 
comparable to the value from Wilson loop. 
With the both fitting functions,
we have obtained the string tension which is comparable to the value from Wilson loop.

This is the first step 
to study the $\bar q$-$q$ potentials from the NBS wave functions,
and the main purpose of the present study is 
to show that the method is applicable to 
the $\bar q$-$q$ potentials.
We find that the obtained $\bar q$-$q$ potential has 
the basic property of that obtained from Wilson loop. 
Therefore, this method can be used for the study of the $\bar q$-$q$ potentials 
with finite quark masses.

\acknowledgments{
The authors 
thank S.~Aoki, T.Doi, T.~Hatsuda, T. Inoue, N. Ishii,
K. Murano, H. Nemura, K. Sasaki,
T.~Kawanai and S.~Sasaki
for the fruitful discussion.
Y.I. also thanks N. Kaiser, A. Laschka and W. Weise for the useful discussion.
The calculations were performed mainly by using the NEC-SX9 and SX8R at Osaka University, and 
partly by RIKEN Integrated Cluster of Clusters (RICC) facility.
This project is supported in part
by Grand-in-Aid for Japan Society for the Promotion of Science (No. 23-8687) and
Scientific Research on Innovative Areas (No. 2004: 20105001, 20105003, 23105713).
}


\appendix
\section{Relativistic effect for the potential}
\label{app:A}

\begin{figure}[t]
\begin{center}
\includegraphics[width=0.85\textwidth]{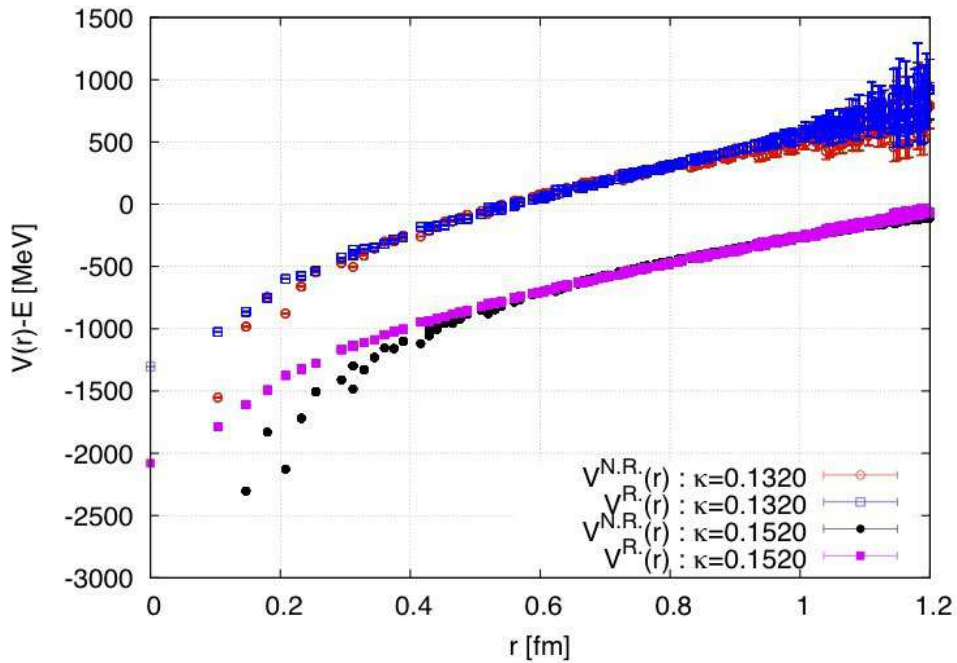}
\caption{Comparison of ``relativistic" potentials $V^{\rm R.}(r)$ 
with non-relativistic potentials $V^{\rm N.R.}(r)$ 
shown in the main part.}
\end{center}
\label{figA4}
\end{figure}
In the main part of the paper, we assume the non-relativistic kinematics for
the Schr\"odinger-type equation. 
We here calculate the potential with 
relativistic kinematics 
to evaluate the relativistic effect for the $\bar q$ - $q$ systems.

The relativistic Schr\"odinger-type equation for the $\bar q$ - $q$ system
in the continuum limit is written as
\begin{eqnarray}
\int d^3 r^\prime\int\frac{d^3p^\prime}{(2\pi)^3}
2\sqrt{\vec p^{\prime 2}+m_q^2}e^{-i\vec p^\prime \cdot (\vec r-\vec r^\prime)}\phi(\vec r^\prime)\nonumber\\
+V(\vec r)\phi(\vec r)=E\phi(\vec r),
\label{rela}
\end{eqnarray}
with the leading order potential $V(\vec r)$ of the velocity expansion.
On a lattice, a discrete Fourier transformation of $\vec r$ 
gives $\sin (\vec p)$. 
Thus $\sqrt{\vec p^{\prime 2}+m_q^2}$ is 
replaced by $\sqrt{(\sin(\vec p^\prime))^2+m_q^2}$, and the 
integral becomes a summation on lattice. 
Similar as the procedure in the main part, 
we obtain the potential from the 
discretized version of the relativistic 
Schr\"{o}dinger-type equation of Eq.~(\ref{rela}).

Fig. (\ref{figA4}) is a comparison of potentials with relativistic kinematics, 
$V^{\rm R.}(r)$ 
and non-relativistic potential $V^{\rm N.R.}(r)$ shown in the main part. 
The relativistic potentials for $\kappa = 0.1320$ ($\kappa = 0.1520$) are shown 
by open square (filled square) in Fig. (\ref{figA4}),
while the non-relativistic potentials for $\kappa = 0.1320$ ($\kappa = 0.1520$) 
are shown by open circle (filled circle).
As shown in Fig. (\ref{figA4}), 
$V^{\rm R.}(r)$ show the linear plus Coulomb behavior, and
$V^{\rm R.}(r)$ and $V^{\rm N.R.}(r)$ is similar for large $r$. 
On the other hand, for small $r$, the difference between them becomes large
as naturally expected, and the short distance differences of the potentials
may contribute to reproducing higher energy states.

We calcurate the root mean square radius $\sqrt{ \langle \vec r^2 \rangle}$
from the NBS wave function
for the heaviest $\bar q$ - $q$ system ($\kappa = 0.1320$) employed in this work.
Then we obtain $\sqrt{ \langle \vec r^2 \rangle} = 0.3550(14)$ fm,
and 
the relativistic potential in Fig. (\ref{figA4}) is consistent with non-relativistic one
around 0.36 fm.
Thus, for the heavy quark case such as the charm quark, we conclude that
the non-relativistic potential picture
gives the proper description of the heavy quarkonium.
On the other hand, for the light quark such as strange quark,
short range part of the relativistic potential differs from non-relativistic one.
This shows that the relativistic effects starts to contribute to
$\bar q$ - $q$ systems.

\section{Potential from a gauge-invariant smeared operator}
\label{app:B}
As we showed, the obtained potentials exhibit a Coulomb plus linear behavior.  
However, potentials with different operators are generally different. 
Therefore, the Cornell-like behavior is not universal. 
Here, we show a potential with a different  operator from that used in the main part.

\begin{figure}[t]
\begin{center}
\includegraphics[width=0.35\textwidth]{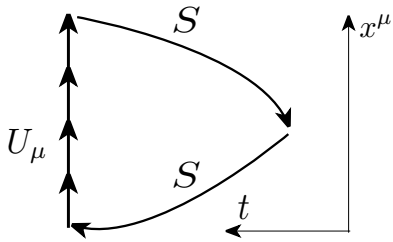}
\end{center}
\vspace{-0.2cm}
\caption{Schematic figure of gauge-invariant smeared operator.
 }
\label{giop}
\end{figure}

The NBS amplitude with gauge-invariant smeared sink operators is defined by 
\begin{align}
&\chi^{\rm smr} (\vec x + \vec r, \vec x, t-t_0; J^{\pi}) \nonumber \\
& \ \ \ \equiv\left\langle 0 \right|
\bar{q}(\vec x + \vec r, t) L(\vec r,\vec x, t;m) \Gamma q(\vec x, t)
\overline{{\cal J}}_{\bar qq}(t_0; J^{\pi})
\left| 0 \right\rangle,\nonumber\\
&L(\vec r=n\hat \mu,\vec x, t)
\nonumber\\
&\equiv U_\mu(\vec x+n\hat\mu,t)\cdots U_\mu(\vec x+\hat\mu,t)U_\mu(\vec x,t) 
\label{gi4-point}
\end{align}
A schematic figure of the amplitude is shown in Fig.~(\ref{giop}). 
The operator $L(\vec r,\vec x, t ; m)$ constructed by link variables connects 
$\vec x$ and $\vec x+ \vec r$ with a straight-line path. 
Here, the direction of $\vec r$ is chosen to be the $x$-, $y$-, or $z$-direction, i.e., on-axes.  

\begin{figure}[h]
\begin{center}
\includegraphics[width=0.80\textwidth]{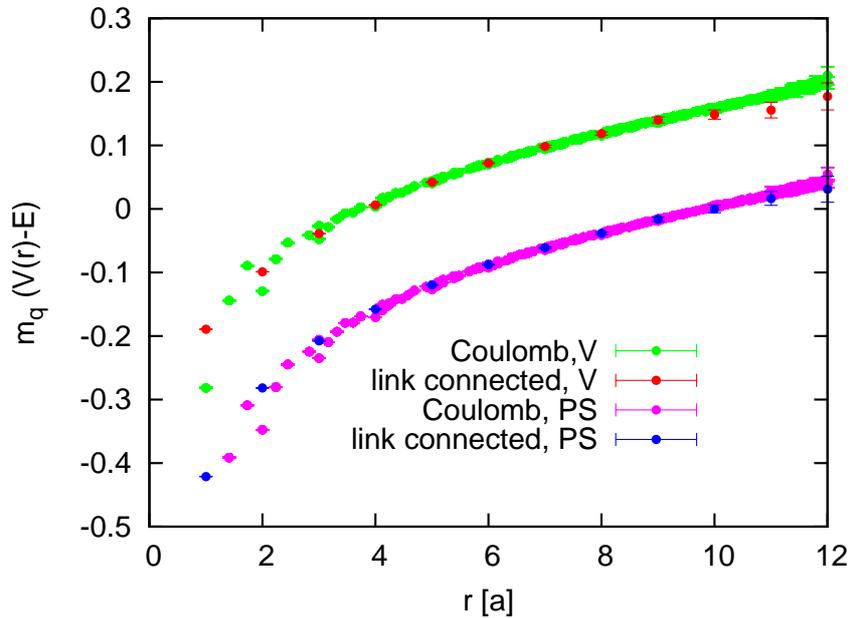}
\end{center}
\caption{
Comparison of a potential with gauge-invariant smeared operator, 
$V^{\rm smr}(r)$, to that with 
Coulomb-gauge operator, $V^{\rm Coul.}(r)$. 
The red (blue) points are 
the data with gauge -invariant smeared operator in V (PS) channel, 
and the green (magenta) points are that in Coulomb gauge in V (PS) channel. 
The two potentials almost coincides.
 }
\label{figB2}
\end{figure}

Figure (\ref{figB2}) shows a potential obtained from the 
smeared NBS amplitude, $V^{\rm smr}(r)$ (red points for V channel and blue points for PS channel), 
and that obtained in Coulomb gauge,
$V^{\rm Coul.}(r)$  (green points for V channel and blue points for PS channel). 
Note that the data of $V^{\rm smr}(r)$ are only calculated on the points with integral multiples 
of the lattice spacing $a$, 
because $\vec r$ in Eq.~(\ref{gi4-point}) is on-axis. 
$V^{\rm smr}(r)$ shows the linear plus Coulomb behavior similar to that in Coulomb gauge, 
and, more over, the two potentials almost coincide. 
This fact shows that the gauge-invariant operator is also a 
suitable one for a constituent quark mass, 
and the Coulomb-gauge operator used in the main part is similar to the gauge-invariant 
operator of Eq.~(\ref{gi4-point}).



\begin{thebibliography}{19}

\bibitem{Chiu1972a}
C.~B. Chiu, 
Ann. Rev. Nucl. Part. Sci. {\bf 22},  255 (1972).




\bibitem{Bali2001}
G.~S. Bali, 
 Phys. Rept. {\bf 343}, 1 (2001).

\bibitem{TTT2000}
  T.~T.~Takahashi, H.~Matsufuru, Y.~Nemoto and H.~Suganuma,
  Phys.\ Rev.\ Lett.\  {\bf 86}, 18 (2001);
  T.~T.~Takahashi, H.~Suganuma, Y.~Nemoto and H.~Matsufuru,
  Phys.\ Rev.\  D {\bf 65}, 114509 (2002).



\bibitem{Brambilla}
  N.~Brambilla, A.~Pineda, J.~Soto and A.~Vairo,
  Nucl.\ Phys.\  B {\bf 566}, 275 (2000);
  Rev.\ Mod.\ Phys.\  {\bf 77}, 1423 (2005).


\bibitem{Brown1979a}
L.~S. Brown and W.~I. Weisberger, 
 Phys. Rev. D {\bf 20}, 3239 (1979).


\bibitem{Eichten1979}
E.~Eichten and F.~L. Feinberg, 
 Phys. Rev. Lett. {\bf 43}, 1205 (1979).

\bibitem{Koma2007a}
Y.~Koma and M.~Koma, 
  Nucl. Phys. {\bf B769}, 79 (2007).

\bibitem{Ishii2007a}
N.~Ishii, S.~Aoki and T.~Hatsuda, 
Phys. Rev. Lett. {\bf 99}, 022001 (2007).

\bibitem{Aoki2010}
S.~Aoki, T.~Hatsuda and N.~Ishii, 
 Prog. Theor. Phys. {\bf 123}, 89 (2010).

\bibitem{Nemura2008}
  H.~Nemura, N.~Ishii, S.~Aoki and T.~Hatsuda,
  Phys.\ Lett.\  \textbf{B673}, 136 (2009).

\bibitem{Nemura2009}
 H.~Nemura {\it et al.} [HAL~QCD and PACS-CS~Collaboration],
 PoS {\bf LATTICE2009}, 152 (2009).

\bibitem{Murano2011}
  K.~Murano, N.~Ishii, S.~Aoki and T.~Hatsuda,
  Prog.\ Theor.\ Phys.\ {\bf 125}, 1225 (2011).
   
\bibitem{Inoue2010a}
  T.~Inoue {\it et al.} [HAL QCD Collaboration],
  Prog.\ Theor.\ Phys.\  {\bf 124 }, 591 (2010).

\bibitem{Inoue2010b}
  T.~Inoue {\it et al.} [HAL QCD Collaboration],
  Phys. Rev. Lett. {\bf 106}, 162002 (2011).

\bibitem{Doi2010}
  T.~Doi {\it et al.} [HAL QCD Collaboration], 
arXiv:1106.2276 [hep-lat];  PoS {\bf LATTICE2010}, 136 (2010).

\bibitem{Sasaki2010}
  K.~Sasaki {\it et al.} [HAL QCD Collaboration],
  PoS {\bf LATTICE2010}, 157(2010);
  arXiv:1012.5684 [nucl-th].

\bibitem{Ikeda2010}
  Y.~Ikeda {\it et al.} [HAL QCD Collaboration],
  arXiv:1002.2309 [hep-lat]; Prog. Theor. Phys. Suppl. {\bf 186}, 228 (2010).

\bibitem{Kawanai2010}
  T.~Kawanai and S.~Sasaki,
  Phys.\ Rev.\  D {\bf 82}, 091501 (2010).


\bibitem{Hatsuda2011}
  T.~Hatsuda,
    arXiv:1101.1463 [nucl-th].

\bibitem{TTT2009}
  T.~T.~Takahashi and Y.~Kanada-En'yo,
  Phys.\ Rev.\  D {\bf 82}, 094506 (2010).

\bibitem{Kroli1956}
W. Kr\'{o}likowski and J. Rzewuski,
Nuovo Cim. {\bf 4}, 1212 (1956).

\bibitem{Levy1952}
M.~M. Levy, 
 Phys. Rev. {\bf 88}, 725 (1952).

\bibitem{Klein1953}
A.~Klein, 
 Phys. Rev. {\bf 90}, 1101 (1953).
\bibitem{Klein1974}
A. Klein and T.-S. H. Lee, Phys. Rev. {\bf 10}, 4308 (1974).

\bibitem{Macke1953}
W. Macke,
Phys. Rev. {\bf 91}, 195 (1953).

\bibitem{IkeIida2010}
  Y.~Ikeda and H.~Iida,
  PoS {\bf LATTICE2010}, 143 (2010).


\bibitem{Iritani2009}
  T.~Iritani, H.~Suganuma and H.~Iida,
  Phys.\ Rev.\  D {\bf 80}, 114505 (2009).

\bibitem{Koma2008}
  M.~Koma, Y.~Koma and H.~Wittig,
  PoS {\bf CONFINEMENT8}, 105 (2008); 
  Y.~Koma and M.~Koma,
  Prog. Theor. Phys. Suppl. No. 186, 205 (2010).

\bibitem{Soto2008}
  G.~Perez-Nadal and J.~Soto,
  Phys.\ Rev.\  D {\bf 79}, 114002 (2009).


\bibitem{Hohler1976}
  G.~Hohler {\it et al.}, 
  Nucl.\ Phys.\  B {\bf 114}, 505 (1976).

\end{thebibliography}

\end{document}